**A defense of using resting state fMRI as null data for estimating false positive rates**

Thomas E. Nichols[a,b], Anders Eklund[c,d,e], Hans Knutsson[c,e]

[a]Department of Statistics, University of Warwick, Coventry CV4 7AL, United Kingdom; and [b]WMG, University of Warwick, Coventry CV4 7AL, United Kingdom; [c]Division of Medical Informatics, Department of Biomedical Engineering, Linköping University, S-581 85 Linköping, Sweden; [d]Division of Statistics and Machine Learning, Department of Computer and Information Science, Linköping University, S-581 83 Linköping, Sweden; [e]Center for Medical Image Science and Visualization, Linköping University, S-581 83 Linköping, Sweden;

ABSTRACT

A recent Editorial by Slotnick (2017) reconsiders the findings of our paper on the accuracy of false positive rate control with cluster inference in fMRI (Eklund et al, 2016), in particular criticising our use of resting state fMRI data as a source for null data in the evaluation of task fMRI methods. We defend this use of resting fMRI data, as while there is much structure in this data, we argue it is representative of task data noise and as such analysis software should be able to accommodate this noise. We also discuss a potential problem with Slotnick's own method.

A recent Editorial by Slotnick (2017) reconsiders the findings of our paper on the accuracy of false positive rate control with cluster inference in fMRI (Eklund et al, 2016). We welcome continued discussion on inferential tools for brain imaging, but felt the need to respond to a number of misrepresentations of our work and highlight potential problems with Slotnick's own method.

The key challenge to our findings is that we made use of resting state fMRI data, which Slotnick asserts invalidates our findings and "provide no basis to question the validity" of fMRI studies. (Please note that our article has been corrected, and now makes no claims about "tens of thousands of published fMRI studies".) However, we carefully considered the use of resting fMRI data, and even dedicated a section entitled "Suitability of Resting-State fMRI as Null Data for Task fMRI", where we enumerated the arguments why this was a reasonable source of null data: (1) While many forms of mental activity may be engaged during rest, only activity synchronized between subjects could give rise to activation in our evaluation; we considered four different types of block and event-related designs, each which resulted in comparable false positive rates for a given smoothing level and analysis tool, suggesting synchronized brain activity was not the source of the inflated false positives. (2) Task residuals have been found to have resting-state networks (Fair et al, 2007), suggesting the spatial covariance structure of fMRI data is similar in task and rest, and we confirmed this by examining spatial autocorrelation function of task residuals and finding them to match the heavy-tailed spatial autocorrelation of the resting data; what's more, this long tailed behavior has also been found in phantom data (Kriegeskorte et al., 2008), indicating that it is intrinsic to the MR acquisition and not even brain-related. (3) We conducted two-sample t-tests on task data, comparing two groups randomly split from a homogeneous group of subjects, and found the same rates of false positive as on pure resting data. This last evaluation is similar to Slotnick (2017), except he constructed null contrasts within subject, then assessed them with a one-sample t-test.

Regarding the location of occurrence of false positive clusters default mode network (DMN), in light of point #2 above, this is a case where resting state "signal" is task fMRI "noise". But this "noise" is highly structured, and characterised by heterogeneous spatial correlation (see Eklund et al, Supplementary Figure 19). We argue that, in order to trust our inferences, our task fMRI methods need to be capable of accommodating of such structured noise.

We also worry that readers will understand the title's "commonly employed methods" to refer to the three major software tools we examined, SPM, FSL & AFNI. Instead, Slotnick (2017) uses a Monte Carlo tool developed in house (`cluster_threshold_beta.m`; downloaded from https://www2.bc.edu/sd-slotnick/scripts.htm 5 Jan. 2017; Slotnick, et al., 2003). Examination of this script shows that it uses Gaussian kernel smoothing of Gaussian noise, as in AFNI's 3dClustSim, though implemented with Fourier convolution. Slotnick (2017) asserts his tool allows that "clusters of activity can have any spatial configuration (i.e., they are not assumed to be Gaussian in shape)"; this is true, but it is equally true for Gaussian random field theory and 3dClustSim; none of these approaches assume Gaussian-shaped clusters, but all of these tools (`cluster_threshold_beta` included) make a very specific assumption that the spatial autocorrelation is proportional to a Gaussian density. Yet it is this particular Gaussian spatial autocorrelation that we found to be inconsistent with real resting and task data, and indeed why it is hard to simulate fMRI data with realistic spatial structure (though see recent work by the AFNI team to improve to 3dClustSim to capture this complexity (Cox et al., 2016)).

Also, while the `cluster_threshold_beta` program makes use of super-resolution for more accurate convolution, it does not appear to correct for edge effects, similar to the bug we found in 3dClustSim. Specifically, Fourier-based smoothing induces dependence between opposing edges that effectively reduces the search space. Hence, `cluster_threshold_beta` too may be providing overly optimistic estimates of familywise error rate, even if no false positives were observed on this one particular dataset examined.

In short, while we acknowledge that yet more evaluations are needed to fully understand the strengths and limitations of task fMRI analysis methods, we assert that large-scale evaluations, not a result from a single dataset as done in Slotnick (2017), are needed to make conclusions about the accuracy of our statistical tools.

REFERENCES

Cox, R. W., Reynolds, R. C., & Taylor, P. A. (2016). AFNI and Clustering: False Positive Rates Redux. *bioRxiv*, 65862. http://doi.org/10.1101/065862

Eklund, A., Nichols, T. E., & Knutsson, H. (2016). Cluster failure: Why fMRI inferences for spatial extent have inflated false-positive rates. *Proceedings of the National Academy of Sciences of the United States of America*, *113*(28), 7900–5. http://doi.org/10.1073/pnas.1602413113

Fair, D. A., Schlaggar, B. L., Cohen, A. L., Miezin, F. M., Dosenbach, N. U., Wenger, K. K., … Petersen, S. E. (2007). A method for using blocked and event-related fMRI data to study "resting state" functional connectivity. *Neuroimage*, *35*(1), 396–405. http://doi.org/10.1016/j.neuroimage.2006.11.051

Kriegeskorte, N., Bodurka, J., & Bandettini, P. (2008). Artifactual time-course correlations in echo-planar fMRI with implications for studies of brain function. *International Journal of Imaging Systems and Technology*, *18*(5–6), 345–349. http://doi.org/10.1002/ima.20166

Slotnick, S. D. (2016). Resting-state fMRI data reflects default network activity rather than null data: A defense of commonly employed methods to correct for multiple comparisons. *Cognitive Neuroscience*, *0*(0), 17588928.2016.1273892. http://doi.org/10.1080/17588928.2016.1273892

Slotnick, S. D., Moo, L. R., Segal, J. B., & Hart, J. (2003). Distinct prefrontal cortex activity associated with item memory and source memory for visual shapes. *Cognitive Brain Research*, *17*(1), 75–82. http://doi.org/10.1016/S0926-6410(03)00082-X